\pdfoutput=1
\documentclass[letterpaper,journal]{IEEEtran}
\usepackage[utf8]{inputenc}
\usepackage{newunicodechar}
\newunicodechar{α}{\ensuremath{\alpha}}
\usepackage{amsmath,amsfonts}
\usepackage{algorithmic}
\usepackage{algorithm}
\usepackage{array}
\usepackage[caption=false,font=normalsize,labelfont=sf,textfont=sf]{subfig}
\usepackage{textcomp}
\usepackage{stfloats}
\usepackage{url}
\usepackage{verbatim}
\usepackage{graphicx}
\usepackage{hyperref}
\usepackage{cite}
\usepackage{booktabs}
\usepackage{multirow}
\usepackage{siunitx}
\usepackage{xspace}

\begin{document}

\title{Audio Physical Dynamics Inspired Deepfake Detection for Voice Authentication Systems}

\author{Alireza Mohammadi, Keshav Sood, Dhananjay Thiruvady, and Asef Nazari
\thanks{The authors are with the School of Information Technology, Deakin University, Geelong, VIC 3220, Australia. \\ (e-mail: alireza.mohammadi9207@gmail.com; keshav.sood@deakin.edu.au; dhananjay.thiruvady@deakin.edu.au; asef.nazari@deakin.edu.au).}}

\markboth{
}
{Mohammadi et al.: Audio Physical Dynamics Inspired Deepfake Detection for Voice Authentication Systems}

\maketitle
\IEEEpubid{\makebox[\columnwidth]{
\hspace{\columnsep}\makebox[\columnwidth]{ }}
\IEEEpubidadjcol
}

\begin{abstract}
Voice authentication systems deployed at the network edge face dual threats: a) sophisticated deepfake synthesis attacks and b) control-plane poisoning in distributed federated learning protocols. 
We present a framework coupling audio physical dynamics deepfake detection with uncertainty-aware in edge learning. The framework fuses interpretable physics features modeling vocal tract dynamics with representations coming from a self-supervised learning module. The representations are then processed via a streamlined Multi-Layer Perceptron backbone, followed by a Bayesian ensemble providing uncertainty estimates. Incorporating audio physical characteristics evaluations and uncertainty estimates of audio samples allows our proposed framework to remain robust to advanced deepfake attacks, while our trust-based aggregation protocol secures the control plane against poisoning in network edge voice authentication systems.
\end{abstract}


\section{Introduction}
\IEEEPARstart{A}{dvanced} neural speech deepfake generation has fundamentally transformed voice authentication security. Modern text-to-speech (TTS) and voice conversion (VC) systems can generate highly realistic audio from minimal target samples, enabling sophisticated attacks ranging from real-time impersonation to automated social engineering at a large scale~\cite{11078629}. This technological advancement has created a critical challenge for voice authentication systems: distinguishing genuine speech from increasingly sophisticated deepfake audio that can convincingly mimic authorized users~\cite{11078629}.

The challenge is compounded by a fundamental limitation in current deepfake detection systems. While state-of-the-art detectors achieve impressive performance on controlled benchmarks with Equal Error Rates (EER) between 0.65\% and 6.56\% on ASVspoof datasets~\cite{WANG2020101114, 10.1109/TASLP.2023.3285283}, their effectiveness degrades substantially against unseen and novel synthesis algorithms. 
This trade-off stems from over-reliance on dataset-specific artifacts rather than fundamental speech production invariants, creating a critical vulnerability as synthesis technology evolves~\cite{WANG2020101114, 10.1109/TASLP.2023.3285283}. 

Contemporary deployment trends introduce a second, equally critical threat. To preserve privacy and reduce latency, modern voice authentication systems increasingly rely on distributed edge learning protocols such as Federated Learning (FL), where models are trained and updated across multiple edge devices\cite{cao2021fltrust}.
However, this distributed approach exposes the system to control-plane attacks: adversarial participants can submit malicious parameter updates designed to compromise the global detection mechanism, even when raw audio never leaves edge devices~\cite{cao2021fltrust}.

Furthermore, existing research in deepfake audio detection has largely addressed these threats in isolation, leaving a critical security gap. The research focuses mainly on improving accuracy against known attacks without modeling distributed deployment risks. Meanwhile, FL security develops domain-agnostic defenses that cannot exploit audio-specific signals or uncertainty quantification. Specifically, a key challenge in FL security is the aggregator's inability to directly inspect raw data on edge devices, making it difficult to distinguish between benign and malicious updates~\cite{Tak2021RawNet2}. This gap between the research focus and FL security direction creates a fundamental vulnerability: no existing framework provides a defense mechanism against both sophisticated deepfake generated audio and distributed learning attacks.\par
We address this gap through a framework that couples audio physical dynamics deepfake detection with Bayesian uncertainty estimate quantification. Specifically, quantifying the uncertainty of the deepfake detection system in its own predictions will serve as a critical metric for assessing trust and screening adversarial client updates in edge learning, without violating the privacy of users. Extensive evaluation on ASVspoof 2019 and 2021~\cite{WANG2020101114, 10.1109/TASLP.2023.3285283} datasets demonstrates the framework's effectiveness across multiple attack scenarios. On logical access (LA) protocols, we achieve 6.80\% EER, with modest degradation on 2021 LA (9.05\% EER) validating cross-dataset generalization. Physical access (PA) evaluation yields 12.95\% EER (2019 PA) and 15.05\% EER (2021 PA), reflecting the increased complexity of replay attack detection across diverse acoustic environments and recording-playback chains.

The contributions of this work are threefold:
\begin{enumerate}
    \item \textbf{Dual-Threat Defense Framework:} simultaneously addressing deepfake voice attacks and control-plane poisoning, bridging a critical gap in existing security research. We propose holistic defense that is more resilient than addressing each threat in isolation.

    \item \textbf{Audio Physical Dynamics and Uncertainty-Aware Detection:} integrates interpretable physics-based speech production features with self-supervised representations, enhanced with Bayesian uncertainty quantification that provides improved detection reliability and verifiable trust signals for distributed deployment.

    \item \textbf{Trust-Based Aggregation Protocol:} A lightweight FL defense for control-plane protection by employing a hybrid anomaly scoring mechanism. By integrating gradient direction, validation loss, and update norms, the protocol screens adversarial client updates to maintain global model integrity against poisoning while preserving user privacy.

\end{enumerate}

\section{RELATED WORK}

Contemporary deepfake detection relies on advanced neural architectures, ranging from end-to-end models like RawNet2~\cite{Tak2021RawNet2} and graph-based AASIST~\cite{Jung2022AASIST} to large-scale self-supervised learning (SSL) systems~\cite{WANG2020101114, 10.1109/TASLP.2023.3285283}. While these detection mechanisms excel at learning discriminative representations from large datasets, their performance degrades when evaluated against unseen deepfake generated samples~\cite{10.1109/TASLP.2023.3285283}.

Physics-guided approaches represent a promising direction for improving robustness. VoiceRadar~\cite{Kumari2025VoiceRadarVD} incorporates interpretable physics-derived features to model micro-frequency dynamics. However, such centralized detectors typically lack calibrated uncertainty quantification and robustness against control-plane attacks in distributed settings.

Moreover, the deployment of machine learning on edge devices via FL introduces critical vulnerabilities. Control-plane poisoning, where adversaries submit malicious parameter updates, remains a persistent threat~\cite{zhou2023securing}. Although defenses like FLTrust~\cite{cao2021fltrust} exist, they often rely on statistical outlier detection that is not suitable to identify domain-specific anomalies in audio representations. 

Our work bridges the gap in mentioned works by coupling audio physical dynamics detection with Bayesian uncertainty. By utilizing calibrated uncertainty for inference reliability and gradient-based screening for training security, the framework addresses the dual threat of deepfake synthesis and control-plane poisoning.

\section{OUR PROPOSED APPROACH}

\subsection{Threat Model}
\label{subsec:threat-model}
We consider an adversary whose objective is to compromise voice authentication system by introducing deepfake speech accepted as genuine. Specifically, the adversary operates across two attack strategies.

\textit{Deepfake attacks.} The adversary generates high-fidelity deepfake voice through two attack vectors: (1) TTS synthesis, where neural vocoders (e.g. WaveNet) generate speech from text input, enabling arbitrary content creation; (2) VC, where source speech is transformed to mimic target speaker characteristics while preserving linguistic content. These attacks constitute the core deepfake threat.

Additionally, the adversary introduces distinct artifacts from recording-playback processes. They execute replay attacks by recording genuine speech of legitimate users and replaying them through various acoustic environments.

\textit{Control-plane attacks.} In distributed edge learning scenarios, the adversary can participate in submitting manipulated parameter updates during aggregation rounds to FL. The manipulations include: (1) \textit{Label-Flip attack}: malicious clients invert labels to corrupt decision boundaries; (2) \textit{Sign-Flip attack}: parameters are negated to ascend the loss surface; (3) \textit{Backdoor attack}: hidden triggers are injected to force targeted misclassification; and (4) \textit{Min-Max attack}: updates are optimized to maximize deviation within Euclidean bounds. We assume a gray-box threat model in which the adversary possesses knowledge of detector architecture and training protocols, but no access to trained model parameters or private training data. The control plane operates under a head-only constraint where only classifier parameters are exchanged.

\subsection{System Architecture}

\begin{figure}[ht!]
    \centering
    \scalebox{1.1667}[1.0]{%
        \includegraphics[width=3in]{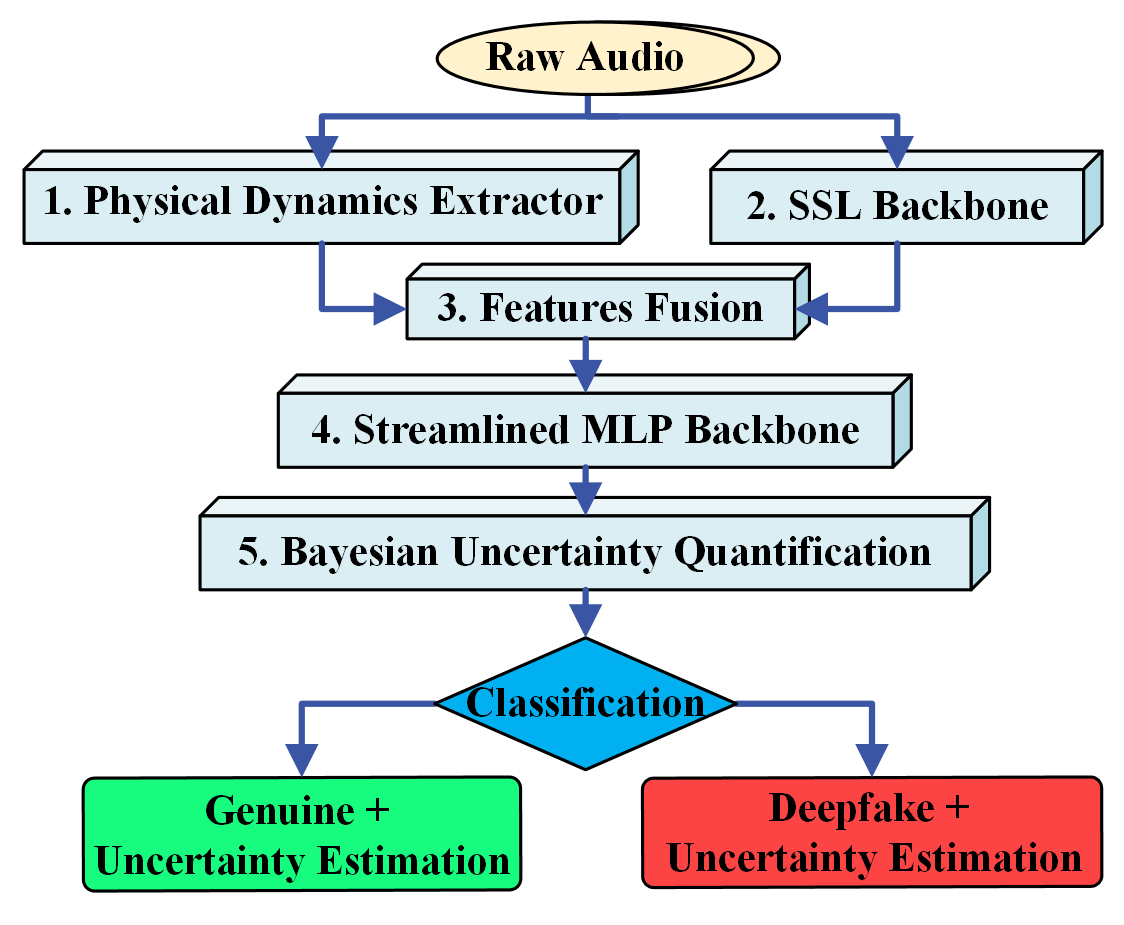}%
    }
    \caption{System architecture overview showing the five-module pipeline for audio physical dynamics deepfake detection. Raw audio (16 kHz, 3-second segments) flows through: (1) Audio physical dynamics Feature Extraction, (2) SSL Backbone (frozen WavLM-Large), (3) Orthogonal Feature Fusion, (4) Bandwidth-Efficient MLP Backbone, and (5) Bayesian uncertainty quantification, producing calibrated genuine/Deepfake audio samples classifications with uncertainty estimates.}
    \label{fig:framework-overview}
\end{figure}

Our framework comprises five modules, summarized in Figure \ref{fig:framework-overview}. Each component contributes a distinct operation in the overall pipeline, and we outline their roles below.

\textit{Module 1: Audio Physical Dynamics Feature Extraction.} This module computes physical representation-dynamics features derived from embedding trajectory analysis. We extract dynamic features modeling translational, rotational, and vibrational patterns. These dynamics-inspired operators serve as interpretable design priors on representation trajectories. Let $\mathbf{E} \in \mathbb{R}^{T \times D}$ denote the SSL embedding sequence of length $T$ with 1024-D, sampled at $f_s = 50$ Hz. We compute velocity $\mathbf{v}_i = (\mathbf{E}_{i+1} - \mathbf{E}_i)/\Delta t$ and acceleration $\mathbf{a}_i = (\mathbf{v}_{i+1} - \mathbf{v}_i)/\Delta t$ at discrete time steps with $\Delta t = 0.02$ s. The translational frequency shift quantifies momentum changes via
\begin{equation}
\Delta f_t = \frac{1}{T-1}\sum_{i=1}^{T-1} \|\mathbf{v}_i\|_2 + \frac{1}{2(T-2)}\sum_{i=1}^{T-2}\|\mathbf{a}_i\|_2,
\end{equation}
where $\|\cdot\|_2$ denotes the $\ell_2$ norm. $\Delta f_t$ averages how fast the embedding moves and how quickly that speed changes. Indexing to $T{-}1$ for $\mathbf{v}_i$ and $T{-}2$ for $\mathbf{a}_i$ reflects finite differences on a length-$T$ sequence, where first and second differences reduce the number of valid points. Vibrational dynamics model oscillatory patterns through power spectral analysis:

\begin{equation}
\Delta f_v = \alpha \cdot \sigma\left( \arg\max_{\omega} \left\{ \left|\mathcal{F}\left(\mathbf{E}_{\cdot,d} \odot w\right)\right|^2\right\}_{d=1}^{D}\right),
\end{equation}

\noindent where $\mathcal{F}$ denotes the discrete Fourier transform, $w \in \mathbb{R}^T$ is a Hann window which tapers sequence endpoints to reduce spectral leakage and stabilize peak estimates, $\odot$ represents element-wise multiplication, $\sigma(\cdot)$ is the standard deviation, and $\alpha = 0.01$ is a normalization constant. Rotational dynamics capture angular momentum via
\begin{equation}
\Delta f_r = \beta \cdot \frac{1}{T-2}\sum_{i=1}^{T-2} \|\mathbf{v}_i \times \Delta\mathbf{v}_i\|_2,
\end{equation}
where $\times$ denotes the cross-product approximation and $\beta = 0.1$ scales to physiological ranges. For PA scenarios, we compute six acoustically-informed features. Dynamic range compression is quantified as
\begin{equation}
R_{\text{dyn}} = 20\log_{10}\left(\frac{\max_t|\mathbf{x}(t)|}{\text{Q}_{0.1}(|\mathbf{x}|)}\right) \text{ dB},
\end{equation}
where $\mathbf{x}(t)$ is the raw waveform and $\text{Q}_{0.1}(\cdot)$ is the 10th percentile. Together, these statistics form a compact 6-D audio physical dynamics vector that summarizes how embeddings move, oscillate, and change direction, and how the audio amplitude is compressed. The vector complementary to the SSL features serves as a discriminative attribute for separating genuine from deepfake audio.

\textit{Module 2: Self-Supervised Learning Backbone.} A \texttt{microsoft/wavlm-large} encoder serves as the SSL backbone, producing 1024-D semantic embeddings that capture high-level statistical patterns in speech.

\textit{Module 3: Orthogonal Feature Fusion.} To ensure complementary rather than redundant information integration, we apply QR decomposition to remove linear dependencies between the $D_{\text{SSL}}=1024$-D SSL embedding and $D_{\text{phys}}=6$-D physics vector. Construct the batch feature matrix
\begin{equation}
\mathbf{X} = \begin{bmatrix} \mathbf{Z}_{\text{SSL}} & \mathbf{Z}_{\text{phys}} \end{bmatrix} \in \mathbb{R}^{B \times (D_{\text{SSL}}+D_{\text{phys}})},
\end{equation}
where $B$ is the batch size, $\mathbf{Z}_{\text{SSL}} \in \mathbb{R}^{B \times 1024}$ contains SSL embeddings, and $\mathbf{Z}_{\text{phys}} \in \mathbb{R}^{B \times 6}$ contains physics features. Center features via $\mathbf{X}_c = \mathbf{X} - \boldsymbol{\mu}$, where $\boldsymbol{\mu} = B^{-1}\sum_{i=1}^B \mathbf{X}_i$ is the feature-wise mean. Then QR decomposition is applied to the transposed centered matrix:
\begin{equation}
\mathbf{X}_c^T = \mathbf{Q}\mathbf{R},
\end{equation}
where $\mathbf{Q} \in \mathbb{R}^{1030 \times 1030}$ is orthogonal. ($\mathbf{Q}^T\mathbf{Q} = \mathbf{I}_{1030}$) 
The orthogonalized features are computed via truncated projection:
\begin{equation}
\mathbf{X}_{\text{ortho}} = \mathbf{X}_c \mathbf{Q}_k \mathbf{Q}_k^T,
\quad 
\end{equation}
where $\mathbf{Q}_k \in \mathbb{R}^{1030 \times k}$ contains the first $k = \min(B, 1030)$ columns of $\mathbf{Q}$. 

The resulting 1030-D orthogonalized representation keeps the 1024-D SSL part and the 6-D audio physical dynamics features linearly decorrelated, preventing the high-dimensional SSL block from overwhelming physics cues. This fused vector is then passed directly to Module 4 as input so both sources influence the downstream decision.

\textit{Module 4: Bandwidth-Efficient Detection Backbone.} The orthogonalized 1030-D feature vector feeds a streamlined Multi-Layer Perceptron (MLP) classifier which transforms them  firstly to 512-D and then to 256-D. This architecture prioritizes efficiency (0.04 ms latency, 660K parameters) over complexity.

\textit{Module 5: Bayesian Uncertainty Decomposition.} This module transforms the embeddings 256-D into calibrated probabilistic outputs with quantified uncertainty. This operation acts as an essential stage for both deepfake classification and uncertainty estimation.
We approximate the posterior distribution over model parameters through stochastic forward passes with dropout enabled during the evaluation phase.

\begin{equation}
\begin{aligned}
\bigl\{\mathbf{p}^{(i)}\bigr\}_{i=1}^{N},\quad
\mathbf{p}^{(i)} &=
\begin{bmatrix}
\hat{p}_{\text{genuine}}^{(i)}\\[2pt]
\hat{p}_{\text{fake}}^{(i)}
\end{bmatrix}\!\in[0,1]^2,\quad
\hat{p}_{\text{genuine}}^{(i)}+\hat{p}_{\text{fake}}^{(i)}=1,\\
\overline{\mathbf{p}} &= \frac{1}{N}\sum_{i=1}^{N}\mathbf{p}^{(i)}.
\end{aligned}
\label{eq:mc-prob-mean}
\end{equation}

Averaging the stochastic samples approximates Bayesian model averaging and provides principled uncertainty quantification.

Standard deterministic classifiers usually rely on outputting only class probabilities without confidence estimates. In edge deployment scenarios where adversarial clients may submit poisoned updates or novel attack types emerge, calibrated uncertainty can enable us to: (1) identify low-confidence predictions requiring human review, (2) screen malicious client updates exhibiting anomalous uncertainty patterns, (3) adapt detection thresholds based on uncertainty. MC dropout enables these capabilities with minimal computational overhead. Additionally, the MLP backbone provides efficient non-linear discrimination of the fused features.


To protect the control plane, we apply a lightweight trust-weighted aggregation on client head updates $\Delta \mathbf{w}_k$. Each update receives a hybrid anomaly score that combines: (1) cosine similarity of gradient direction to a benign prior, (2) validation loss on a small server-held probe set, and (3) deviation in update norm to reveal scaling attacks. The system rejects low-trust score updates before aggregation, preserving global model integrity against control plane poisoning while keeping client data private.

\section{EXPERIMENTAL EVALUATION \& DISCUSSION}
\subsection{Datasets and Evaluation Protocols}
We conduct a comprehensive evaluation using the ASVspoof benchmark datasets\cite{WANG2020101114, 10.1109/TASLP.2023.3285283}. The primary evaluation employs ASVspoof 2019 LA and PA protocols, containing 25,380/24,844/71,237 and 54,000/24,300/137,538 samples in train/development/test sets, respectively. Both protocols exhibit severe class imbalance with samples labeled as genuine comprising approximately 10\% of each set and around 90\% of samples labeled as deepfake generated (TTS and VC version for LA dataset). For cross-dataset assessment, we evaluate our model's performance which has been trained on ASVspoof 2019 LA and PA against unseen ASVspoof 2021 LA and PA sets respectively. The 2021 LA test set includes 148,000 deepfake audio samples, while the 2021 PA set contains 137,538 samples. \par 
All audio samples undergoes consistent preprocessing: resampling to 16 kHz, segmentation into 3-second non-overlapping windows, and normalization to [-1, 1] dynamic range. System performance is quantified using three complementary metrics: EER, ROC-AUC, and min-tDCF. Edge deployment analysis on Apple M4 Pro hardware (8-core CPU, 48 GB RAM) demonstrates practical feasibility: 149 ms end-to-end latency per 3-second audio segment, equivalent to 6.7 files/second throughput. The frozen SSL module's backbone model requires 1.26 GB storage and 2.31 GB inference memory compatible with commodity edge servers and high-end mobile devices (flagship smartphones, IoT gateways). The performance can be feasible in deployment for batch audio analysis applications (voicemail screening, recorded call authentication).
Software implementation employed AI-assisted development tools (Cursor IDE with Claude 4.1 and GPT-5) for the following purposes: systematic debugging and validation, code refactoring, and workflow documentation.

\subsection{Detection Performance}


Table~\ref{tab:main-results} presents our framework's detection performance across ASVspoof 2019 and 2021~\cite{WANG2020101114, 10.1109/TASLP.2023.3285283} LA and PA protocols. On ASVspoof 2019 LA, we achieve 6.80\% EER with min-tDCF of 0.0187 and ROC-AUC of 0.9751, demonstrating low detection cost at optimal thresholds and strong class separability. Cross-dataset evaluation on ASVspoof 2021 LA yields 9.05\% EER (min-tDCF: 0.0155, ROC-AUC: 0.9785). 
Physical access protocols show 12.95\% EER (2019 PA) and 15.05\% EER (2021 PA), with min-tDCF values of 0.0383 and 0.0427 maintaining application-feasible detection costs despite increased acoustic complexity. 

\begin{table}[h]
\centering
\caption{Detection Performance Across ASVspoof Protocols (Robust Framework)}
\label{tab:main-results}
\scriptsize
\begin{tabular*}{\columnwidth}{@{\extracolsep{\fill}}l c c c@{}}
\toprule
\textbf{Protocol} & \textbf{EER (\%)} & \textbf{min-tDCF} & \textbf{ROC-AUC} \\
\midrule
ASVspoof 2019 LA & 6.80 & 0.0187 & 0.9751 \\
ASVspoof 2021 LA & 9.05 & 0.0155 & 0.9785 \\
ASVspoof 2019 PA & 12.95 & 0.0383 & 0.9332 \\
ASVspoof 2021 PA & 15.05 & 0.0427 & 0.9223 \\
\bottomrule
\end{tabular*}
\end{table}

Our detection backbone achieves 0.23\% EER on ASVspoof 2019 LA, competitive with VoiceRadar~\cite{Kumari2025VoiceRadarVD} (0.10\% EER), AASIST~\cite{Jung2022AASIST} (0.83\% EER), and substantially outperforming RawNet2~\cite{Tak2021RawNet2} (5.13\% EER) and LCNN~\cite{Lavrentyeva2019LCNN} (5.06\% EER). The 0.13 percentage point gap to VoiceRadar enables calibrated uncertainty quantification and control-plane defense, capabilities absent in centralized detectors.



\begin{figure}[h]
\centering
\begin{tabular}{@{}c@{\hspace{2pt}}c@{}}
\includegraphics[height=3.22cm]{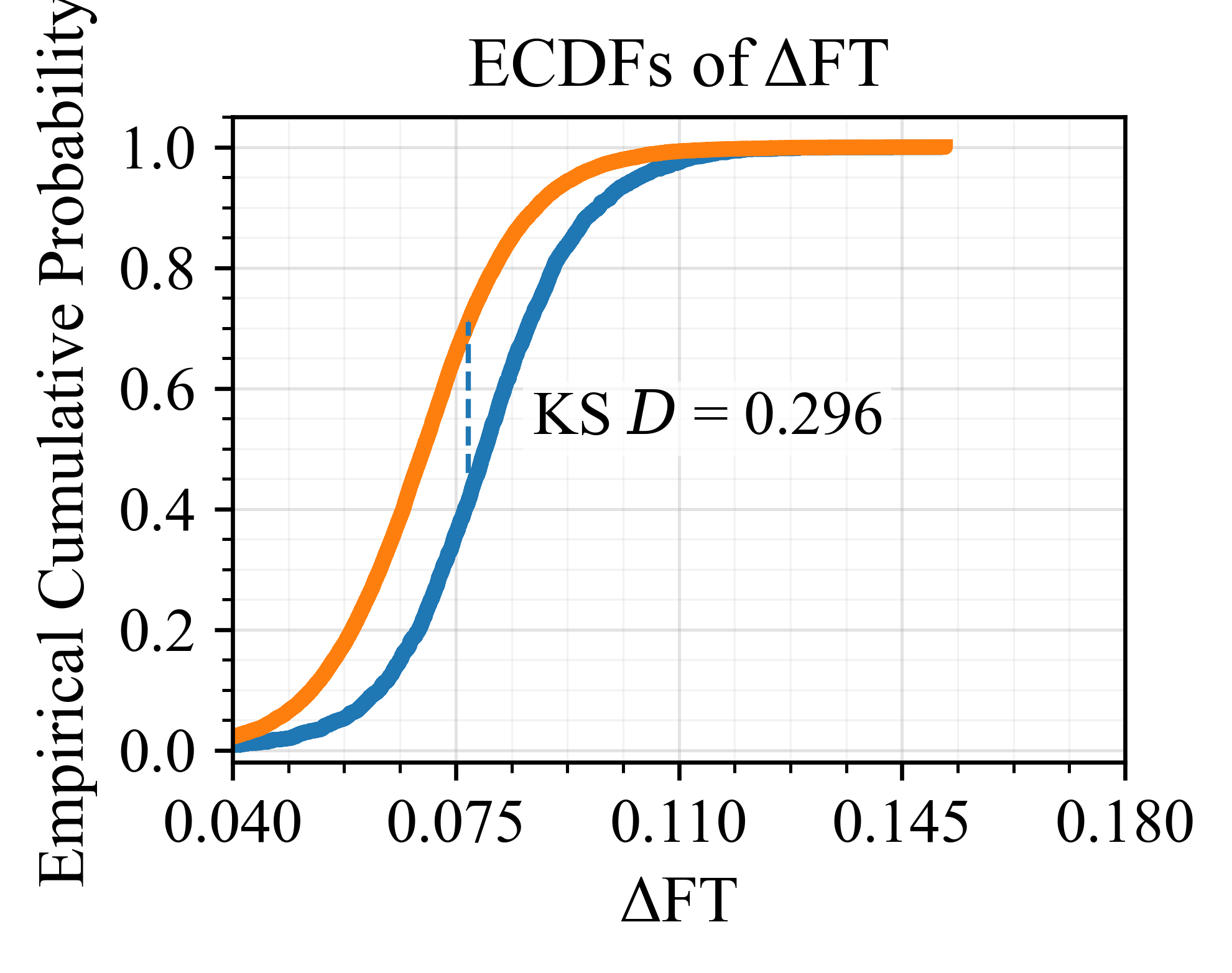} &
\includegraphics[height=3.22cm]{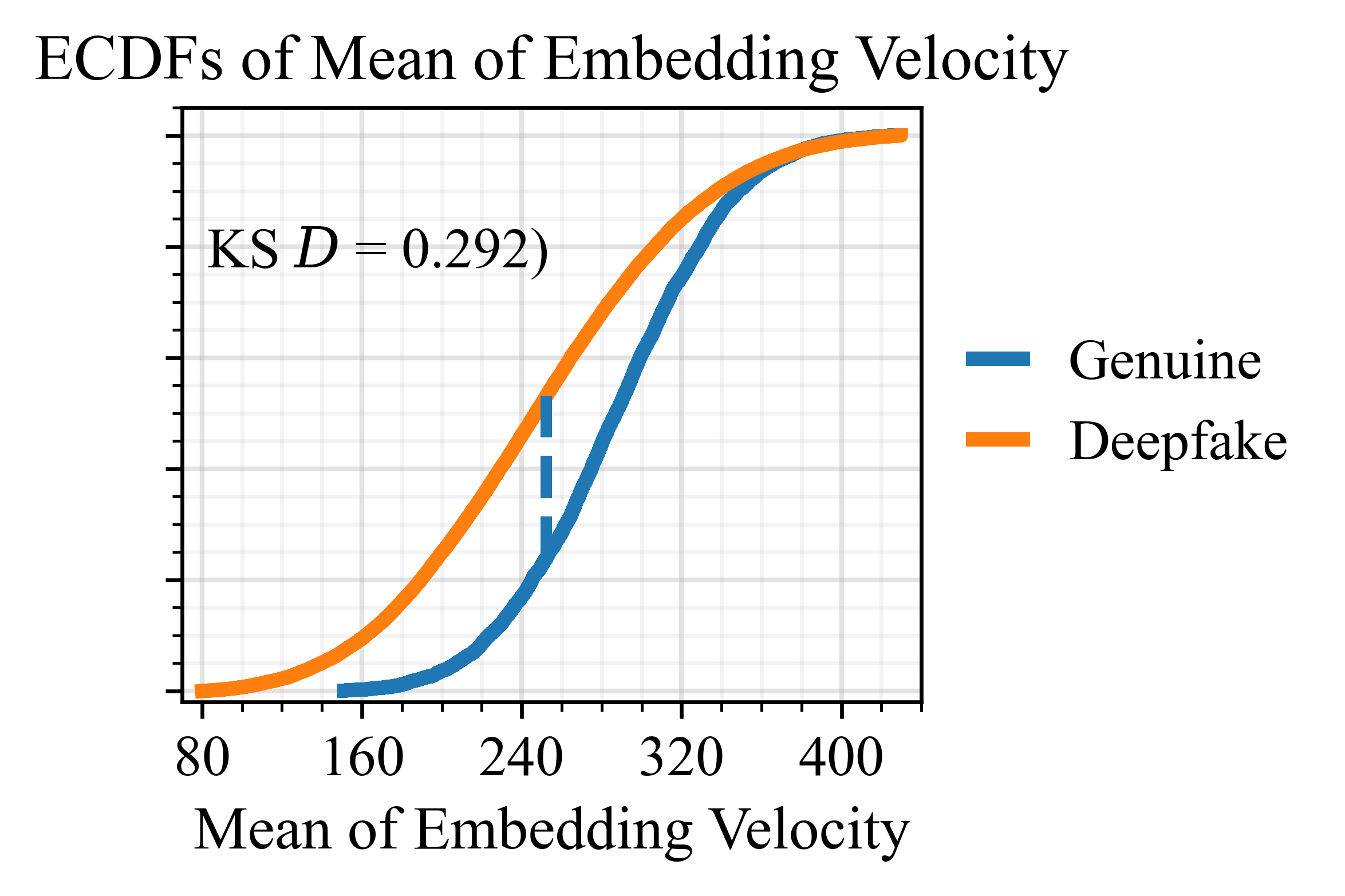} \\
(a) & (b)
\end{tabular}
\caption{Empirical cumulative distribution functions on ASVspoof 2019. (a) Temporal-frequency variation ($\Delta$FT) for LA, where the deepfake distribution shifts left relative to genuine speech, yielding KS distance $D \approx 0.296$ and ROC-AUC $= 0.697$. (b) Embedding mean velocity magnitude for PA, where a similar leftward shift produces KS distance $D \approx 0.292$, confirming discriminative separation between genuine and deepfake speech.}
\label{fig:ecdf}
\end{figure}

We validate the fundamental premise that audio physical dynamics features provide discriminative signals through empirical cumulative distribution function (ECDF) analysis on training partitions.
ASVspoof LA' evaluation scenarios demonstrate similar discriminative patterns through the \texttt{temporal-frequency variation} feature (Fig.~\ref{fig:ecdf}(a)). Deepfake audio samples show systematic leftward distribution shift with median reduction from 0.0847 to 0.0655, corresponding to $D = 0.296$ and ROC-AUC = 0.697. This statistical attribute validates that deepfake generated samples introduce temporal-frequency artifacts detectable through physics-informed feature engineering.\par 
For PA scenarios, the \texttt{embedding mean velocity magnitude} exhibits class separation (Fig.~\ref{fig:ecdf}(b)). 
Deepfake generated audio distributions shift leftward with mean values dropping from 288.04 to 247.31, yielding Kolmogorov-Smirnov distance $D = 0.292$ and univariate ROC-AUC of 0.683. This separation reflects the audio physical dynamics-based hypothesis that replay attacks introduce velocity artifacts inconsistent with natural speech production dynamics.

Sensitivity analysis confirms robustness to hyperparameter choice. Varying scaling constants $\alpha$/$\beta$ across a $3 \times 3$ grid yields minimal EER variation (0.16\%--0.27\%). Furthermore, dynamics features consistently improve the WavLM backbone ($\Delta \text{EER} = -0.06$\%), suggesting a specific synergy with WavLM's representation space.

\subsection{Architecture Selection and Ablation.}

We validate our architectural selection via a unified ablation study on the ASVspoof 2019-LA development partition (Table~\ref{tab:ablation-dev}). Results decisively favor the streamlined MLP backbone (0.16\% EER) over both the complex ViT+GNN composite (0.22\% EER) and the SSL-only baseline (0.22\% EER). The integration of orthogonalized dynamics features yields a net improvement of $\Delta \text{EER} = -0.06$\%. Moreover, the MLP architecture reduces parameter overhead by 4.6\% compared to attention-based variants while preserving negligible inference latency (0.04 ms).

\begin{table}[h]
\centering
\caption{Architecture Ablation (ASVspoof2019-LA, 5-fold CV)}
\label{tab:ablation-dev}
\scriptsize
\begin{tabular*}{\columnwidth}{@{\extracolsep{\fill}}l c c c@{}}
\toprule
\textbf{Configuration} & \textbf{EER (\%)} & \textbf{Params} & \textbf{Latency (ms)} \\
\midrule
SSL + Dynamics (MLP) & \textbf{$0.16\!\pm\!0.05$} & 660K & 0.04 \\
SSL-only & $0.22\!\pm\!0.06$ & 657K & 0.04 \\
(ViT+GNN) & $0.22\!\pm\!0.09$ & 692K & 0.05 \\
Linear & $0.28\!\pm\!0.08$ & 2{,}062 & 0.03 \\
Audio Dynamics-only & $15.09\!\pm\!0.40$ & 14 & 0.03 \\
\bottomrule
\end{tabular*}
\end{table}


We validate the probabilistic reliability of our framework on the LA-eval partition. 

The Bayesian MC-Dropout ensemble ($T=30$) improves Expected Calibration Error (ECE) from 0.0007 (deterministic baseline) to 0.0004, achieving near-perfect fidelity without compromising discriminative power (EER remains 0.23\%). This calibration enables effective inference-time screening: rejecting the top 5\% most uncertain predictions (95\% coverage) eliminates false acceptances completely, reducing the FAR from 0.04\% to 0.00\%. These results confirm that quantified uncertainty serves as a decisive proxy for error potential, providing a tangible security advantage in high-stakes deployment.


\subsection{Federated Control-Plane Robustness}

We evaluate trust-weighted aggregation with $N{=}50$ clients, 20\% round participation, and non-IID partitioning on frozen WavLM embeddings. We compare five baselines across six scenarios: clean, label-flip, sign-flip (defended), plus Gaussian noise, Min-Max, backdoor (stress tests). Table~\ref{tab:fl-full} reports mean EER over three seeds.

Trust-Weighted holds 2.06\% EER under clean FL and degrades minimally to 2.57\% (label-flip) and 2.74\% (sign-flip) at $\rho{=}20\%$. FedAvg~\cite{pmlr-v54-mcmahan17a} drops from 2.75\% to 4.72\% under sign-flip, confirming attack strength. Under Gaussian noise, Median~\cite{pmlr-v80-yin18a} achieves 3.44\% EER via coordinate-wise filtering, while our trust scoring yields 5.40\%. Min-Max severely impacts Krum~\cite{GARCIAMARQUEZ2025114004} (39.36\% EER); Trust-Weighted sustains 2.38\%. For backdoor, we recover 2.01\% EER. Our gradient-based trust mechanism screens label and gradient manipulation, providing control-plane security in head-only FL.



\begin{table}[h]
\centering
\caption{FL Robustness: EER (\%) Under All Attack Scenarios}
\label{tab:fl-full}
\scriptsize
\setlength{\tabcolsep}{1.8pt}
\renewcommand{\arraystretch}{0.96}
\begin{tabular}{@{}l c c c c c c@{}}
\toprule
\textbf{Attack} & \textbf{FedAvg}~\cite{pmlr-v54-mcmahan17a} & \textbf{Median}~\cite{pmlr-v80-yin18a} & \textbf{Trim}~\cite{pmlr-v80-yin18a} & \textbf{Krum}~\cite{GARCIAMARQUEZ2025114004} & \textbf{FLTrust}~\cite{cao2021fltrust} & \textbf{Ours} \\
\midrule
Clean        & 2.75$\pm$2.0 & 2.55$\pm$0.3 & 2.35$\pm$0.2 & 6.13$\pm$1.4  & 10.39$\pm$1.2 & 2.06$\pm$0.2 \\
Label-flip & 3.33$\pm$1.9 & 3.22$\pm$0.6 & 3.18$\pm$0.4 & 12.41$\pm$4.1 & 10.57$\pm$1.1 & 2.57$\pm$0.4 \\
Sign-flip  & 4.72$\pm$3.7 & 3.22$\pm$0.4 & 3.16$\pm$0.2 & 9.14$\pm$3.6  & 10.99$\pm$1.5 & 2.74$\pm$0.3 \\
\midrule
Gaussian   & 4.52$\pm$0.3 & 3.44$\pm$0.3 & 4.58$\pm$0.2 & 6.02$\pm$1.0  & 11.13$\pm$1.2 & 5.40$\pm$0.2 \\
Min-Max   & 2.73$\pm$1.7 & 3.44$\pm$0.4 & 2.90$\pm$0.3 & 39.36$\pm$32.4 & 11.08$\pm$2.4 & 2.38$\pm$0.2 \\
Backdoor & 2.85$\pm$2.1 & 2.86$\pm$0.7 & 2.38$\pm$0.5 & 8.17$\pm$2.8  & 10.57$\pm$1.1 & 2.01$\pm$0.4 \\
\bottomrule
\end{tabular}
\end{table}

Furthermore, we evaluate the ability of the trust mechanism to distinguish benign from malicious clients using trust scores averaged across rounds (Table~\ref{tab:trust-diagnostics}). Under label-flip and sign-flip attacks, the trust scores provide measurable separation, with AUROC values of 0.853 and 0.894 and corresponding True Detection Rates (TDR) of 43.1\% and 49.6\%. Despite weaker separability detection under Min-Max, Trust-Weighted aggregation maintains the lowest EER (2.38\%), indicating robustness to malicious-client identification. 

\begin{table}[h]
\centering
\caption{Trust Diagnostics at 5\% FDR Threshold}
\label{tab:trust-diagnostics}
\scriptsize
\begin{tabular*}{\columnwidth}{@{\extracolsep{\fill}}l c c c c c@{}}
\toprule
\textbf{Attack} & \textbf{Benign} & \textbf{Malicious} & \textbf{Gap} & \textbf{TDR@5\%} & \textbf{AUROC} \\
\midrule
Gaussian & 0.760 & 0.545 & 0.215 & 58.2\% & 0.935 \\
Label-flip & 0.708 & 0.586 & 0.122 & 43.1\% & 0.853 \\
Sign-flip & 0.702 & 0.550 & 0.152 & 49.6\% & 0.894 \\
Min-Max & 0.669 & 0.582 & 0.088 & 37.9\% & 0.755 \\
Backdoor & 0.680 & 0.664 & 0.016 & 3.4\% & 0.565 \\
\bottomrule
\end{tabular*}
\end{table}

\section{CONCLUSION}
In this work, we addressed the dual threat of deepfake attacks and control-plane poisoning in distributed voice authentication systems. We introduced a framework that integrates audio physical dynamics detection with Bayesian uncertainty quantification. Our results demonstrated that audio physical dynamics-informed features provide consistent discriminative power between deepfake generated and genuine audio samples. Furthermore, our trust-weighted aggregation protocol successfully neutralizes Label-Flip and Sign-Flip poisoning attacks in head-only federated learning, maintaining performance comparable to the clean baseline. While the system effectively secures standard distributed deployment, future work entails enhancing resilience against adaptive optimization-based attacks and exploring defense distillation for extreme edge constraints.

\bibliographystyle{IEEEtran}
\bibliography{reference}

\vfill
\end{document}